\documentclass[boxit]{JAC2003}
\usepackage{graphicx}

\begin{document}
\title{SYNCHROTRON RADIATION INTERFEROMETER CALIBRATION CHECK BY USE OF A SIZE CONTROL BUMP IN KEKB}
\author{N. Iida, J. Flanagan, Y. Funakoshi and K. Oide,
 \\KEK, 1-1 Oho, Tsukuba, Ibaraki, 305-0801, Japan}

\maketitle

\begin{abstract}

In KEKB, synchrotron radiation interferometers (SRMs)\cite{SRM} are used for measuring the transverse beam sizes.
There is also a tool for enlarging the vertical beam size intentionally by making an asymmetric bump, called an ``iSize'' bump, at  one of the strongest non-interleaved  sextupole magnets in each KEKB ring.
The calibrations of the SRMs were checked by comparing the measured vertical beam sizes with those calculated using the computer code ``SAD''. 
The obtained correction factors are 1.000$\pm$0.045 for HER and 0.971$\pm$0.060 for LER, which are  consistent with the calibration factors of SRMs\cite{SRMcalib} within errors. 
Using the obtained calibration factor,  x-y coupling of each ring was calculated .
\end{abstract}

\section{INTRODUCTION}

KEKB is a colliding-beam accelerator aimed at high luminosity.
There are two rings in KEKB, one called the High Energy Ring (HER) and the other called the Low Energy Ring (LER), whose energies are 8.0 GeV and 3.5 GeV, respectively.
Beam-beam simulations show that it is necessary to keep single-beam vertical size small for obtaining a high luminosity.
Therefore it is essentially important to measure the absolute values of the vertical beam sizes.
In addition x-y coupling is one of the most important parameters to diagnose optics of a machine, 
which can be estimated by measuring the absolute value of beam size.
The calibration factor of the SRM is obtained by making a local bump at the luminous point\cite{SRMcalib}. 
Here we measured it by a quite different method in which a size control bump, called ``iSize'' bump, is used. 
The iSize bump in HER was used as a feedback knob for keeping a vertical size ratio of the electron beam to the positron constant during the collision.
It  was useful to get high luminosity at early KEKB operation in about 2000\cite{iSizeFB}.

\section{iSize BUMP}

Both of the KEKB rings have four straight sections and four arcs. 
At KEKB a non-interleaved sextupole scheme is adopted to correct chromaticity. 
Therefore many pairs of sextupole magnets are installed  in the arc sections where the large horizontal dispersions exist.
At one of the strongest sextupole pairs in a ring, an anti-symmetric bump (iSize bump) is made by using three dipole correction magnets near them.
As the phase advance between the pair of sextupole magnet is $\pi$, the x-y coupling is confined to the bump and does not leak out. 
This x-y coupling converts the horizontal dispersion to the vertical which leaks out around the whole of the ring.
At the arc section the beam emits synchrotron radiation light,
which causes the radiation excitation in the place where the vertical dispersion exists, and consequently enlarges the vertical beam emittance and beam size.  

\begin{figure}[hbt]
\centering
\includegraphics*[width=82.5mm]{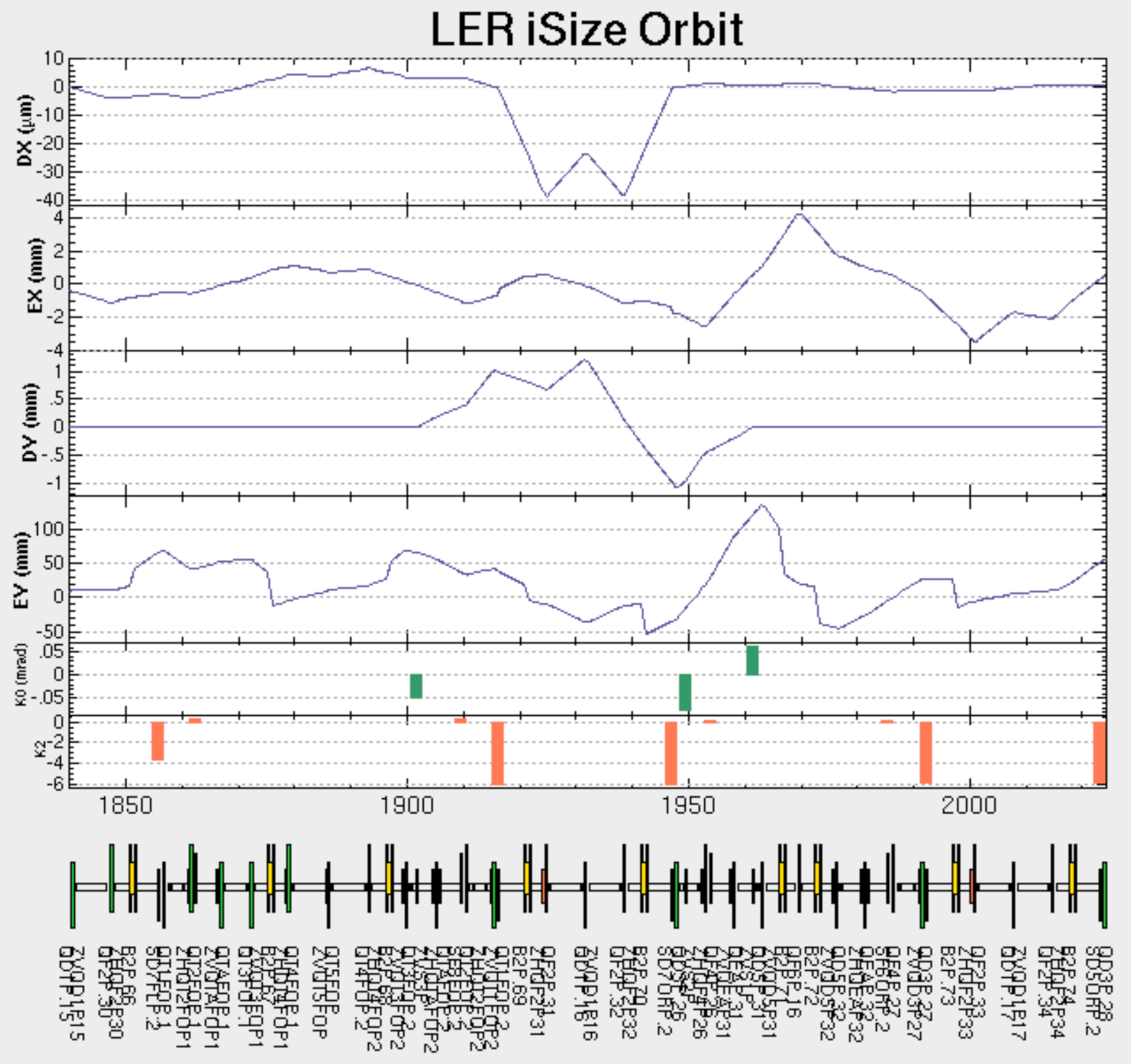}
\caption{LER orbits calculated with the SAD code.
 From the top, the horizontal orbit, dispersion, the vertical orbit, dispersion, strengths of correction magnets and those of sextupole magnets are indicated.}
\label{Orbit}
\end{figure}

The calculated LER orbits and dispersions around the pair of sextupoles with a bump of 1 mm height at one of the sextupole pairs is shown in Fig.~\ref{Orbit}. 
A large leakage of the vertical dispersion out of the bump can be observed.
In actual operation, a continuous closed orbit correction(CCC) system flattens the orbit with reference to the optimized (``golden'') orbit, every 20 seconds.
When an iSize bump is made, CCC adds the iSize bump pattern and its height to the golden orbit.

The variation of the vertical beam size by changing the iSize bump height can be represented as:
\begin{eqnarray}
\sigma_Y^2&=&\sigma_{Y0}^2+A^2(h-h_0)^2 \nonumber\\
(\sigma_Y^{meas})^2&=&(c\cdot\sigma_Y)^2\nonumber\\
&=&(c\cdot\sigma_{Y0})^2+(cA)^2(h-h_0)^2,\label{sigma2}
\end{eqnarray}
where $\sigma_Y$, $\sigma_{Y0}$ and $\sigma_Y^{meas}$ are the true vertical beam size with an iSize bump, that without iSize bump and that measured by SRM, respectively. 
An iSize bump height and its offset are indicated by $h$ and $h_0$ which is necessary for fitting the measured data as described later.
$A$ is a linear coefficient of the vertical beam size {\it vs.} iSize bump height, which is a parameter determined by the optics model.
The calibration factor for the SRM, $c$, can be obtained from $cA$ measured by SRM divided by $A$ calculated by simulation.

\section{MEASUREMENTS}

\begin{figure}[htbp]
\centering
\includegraphics*[width=82.5mm,]{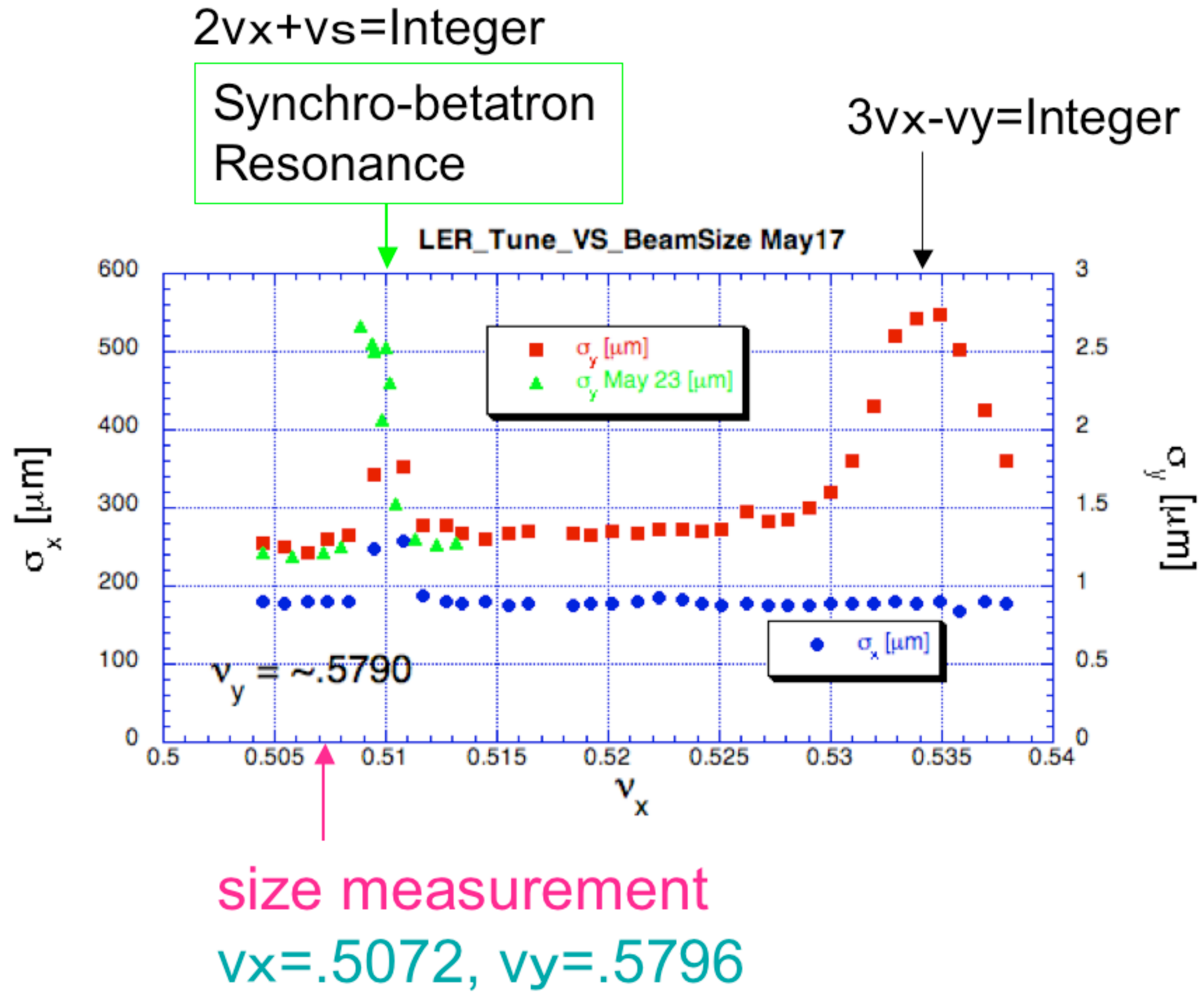}
\caption{Measured beam size dependence on the horizontal tunes in LER.
	The blue dots indicate the horizontal beam sizes and the red and green dots are the vertical beam sizes.
	These beam sizes were scaled to those at the interaction point (IP) using the optics model.
	 }
\label{Tune}
\end{figure}

There are some resonance excitations and growths of beam sizes, which should be avoided.
The resonances that we were primary concerned with were 2$\nu_X+\nu_S$=Integer and 3$\nu_X-\nu_Y$=Integer, where $\nu_X$ and $\nu_S$ is the horizontal betatron tune and synchrotron tune, respectively.
The values of $\nu_S$ are -0.0022 in both rings.
Just before measuring beam sizes by making iSize bumps, we measured the dependence of the transverse beam size on the horizontal tune in each ring.
The two resonances of the beam sizes are observed as large peaks near the operating tune area in the LER, shown in Fig.~\ref{Tune}.
We chose 0.5072 as the fractional part of the horizontal betatron tune.
The fractional part of the vertical betatron tune of the LER was set to 0.5796 in the measurements during the tune search and the vertical beam size measurements with iSize bumps.
The tunes were also used in KEKB operation.
The tune feedback system always maintained the measured tunes back at the set values throughout the iSize measurement.

\begin{figure}[htbp]
\centering
\includegraphics*[width=68.5mm]{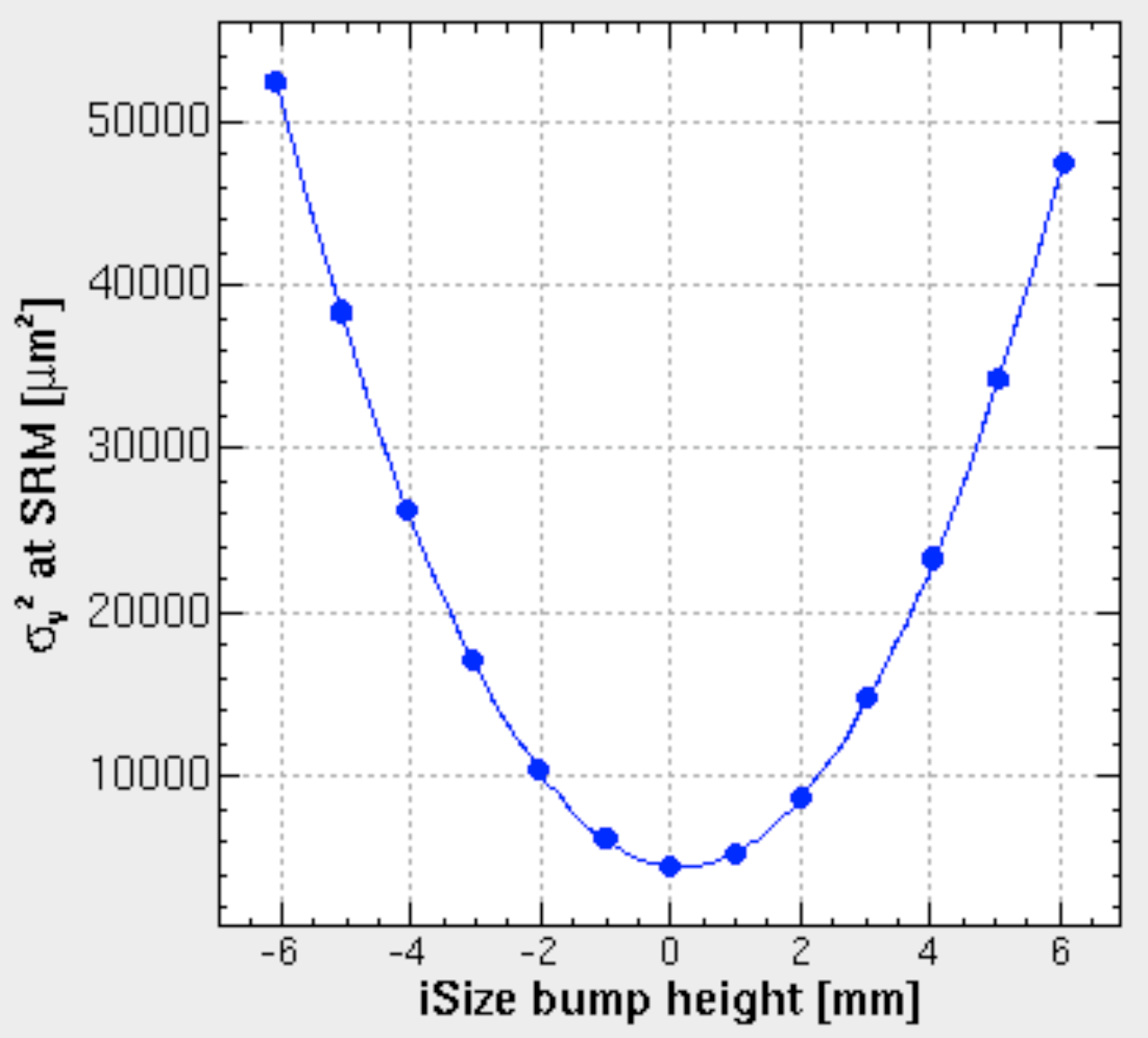}
\caption{The vertical beam sizes measured by SRM versus iSize bump height in the LER are shown as blue dots.
	The blue line fits the points by the function of Eq.~(\ref{sigma2}).}
\label{Parabola}
\end{figure}

The squares of the measured vertical beam sizes as functions of iSize bump heights in the LER are shown in Fig.~\ref{Parabola}. 
At each step we took data for one minute, starting one minute after making an iSize bump, to wait for the CCC to flatten the orbit, and to let the moving average SRM data settle (decay time of about 20 seconds).
The bottom of the parabolic fit in Fig.~\ref{Parabola} is shifted $h_0$ in Eq.~\ref{sigma2} due to the original vertical dispersion at the iSize bump.
The values of $h_0$ in Eq.~(\ref{sigma2}) are 0.177 mm and 0.160 mm in the HER and the LER, respectively.
The amplitude of the minimum beam size is caused by the original vertical emittance due to machine errors and the vertical betatron oscillation from x-y coupling.

\begin{figure}[htbp]
\centering
\includegraphics*[width=70.5mm]{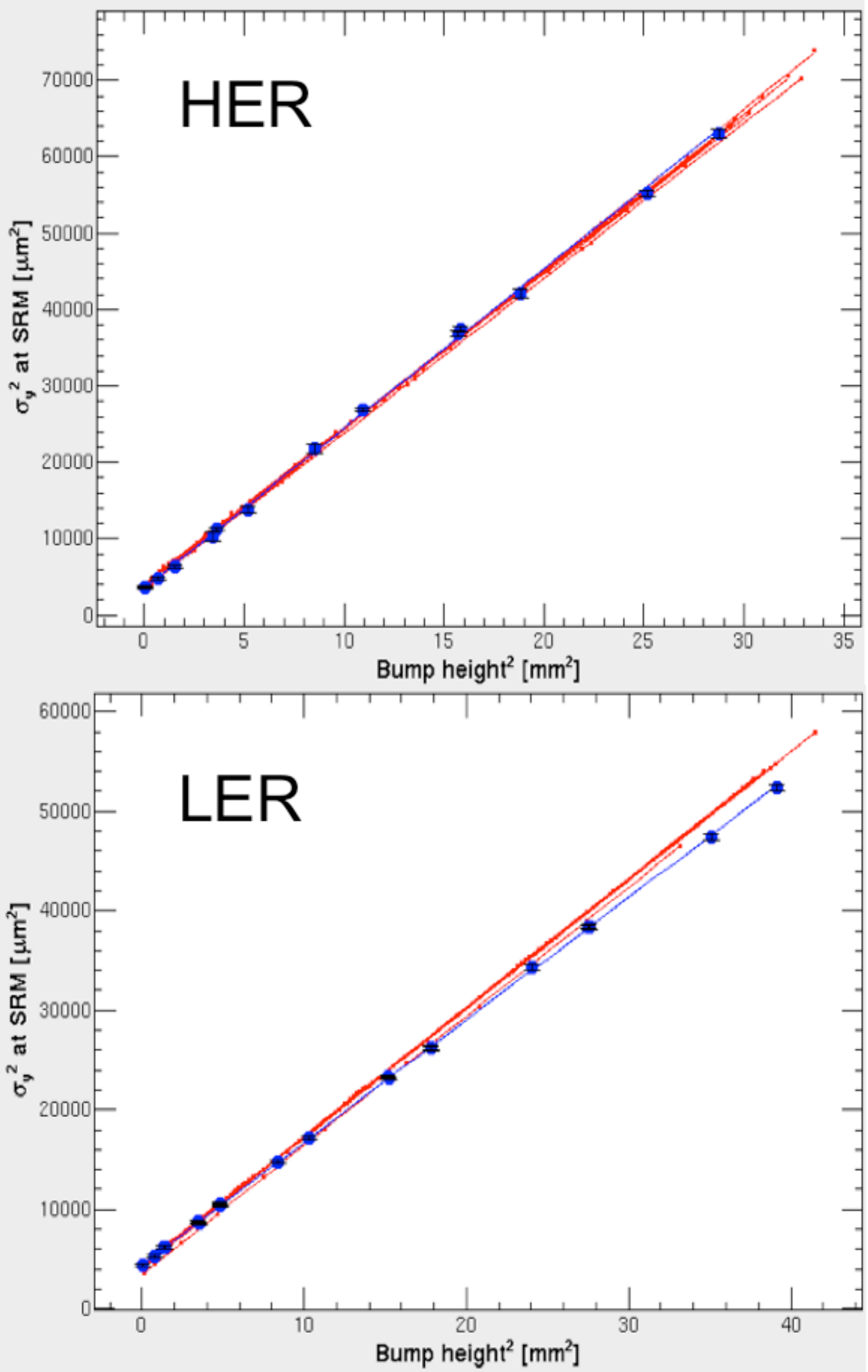}
\caption{The squares of the vertical beam sizes are plotted against the squares of iSize bump heights.
The blue line indicates measurements with the SRM, and the red lines indicate calculations by SAD.
}
\label{Result}
\end{figure}

The squares of the vertical beam sizes {\it vs.} the squares of the iSize bump heights are shown in Fig.~\ref{Result}.
The blue points are the same values as in Fig.~\ref{Parabola},
fit very well to Eq.~(\ref{sigma2}), as shown by the blue lines.

\section{SIMULATION}\
All optics calculations are done with the ``SAD'' code~\cite{SAD} in which a lattice for the current operations are used~\cite{SADope}.
SAD calculates the equilibrium 6D beam envelope all around the ring  
to estimate the beam size,  by taking the radiation damping and the  
excitation into account. All effects from the x-y coupling and the  
dispersion are automatically included.
Details are described in \cite{size}.
We performed simulations just as done in the measurement, that is, after setting iSize bump, the tunes were set to the chosen values described above, by moving strength values of a number of particular  quadrupole magnets in the ring.
Just before this measurements, the global betatron function in each ring was measured.
The ratios of the measured square roots of betatron functions to the design values at the SR source points are 0.985 and 1.00, for the HER and the LER, respectively.
Corrections were done by using these ratios to the beam size simulations.
The vertical beam size without iSize bump cannot be determined only from simulation.
We simulated the beam size by giving vertical offsets to all sextupole magnets in the ring.
The offset values were chosen to agree with measurement within $\pm$10 \%.
The offset at each sextupole magnet was set by generating random numbers distributed as a Gaussian.
Ten combinations of random offsets at all sextupoles were simulated and the results are shown in Fig.~\ref{Result} as ten red lines.
The iSize bump heights were also shifted by the height where the parabola bottoms exist as same as done in measurements. 
The slopes of the ten lines are in good agreement with each other, which means that the dependence of the vertical beam size on iSize bump height does not depend on the vertical offsets of the sextupole magnets. 

\section{CALIBRATION FACTOR}\
We obtained calibration factors for the SRMs of 1.000$\pm$0.006 and 0.971$\pm$0.001 in the HER and the LER, respectively, from Fig.~\ref{Result}.
The errors are from statistic errors in the measurements and the simulations.
Systematic errors are considered as follows;
(1) Uncertainties of about $\pm$4 \% and $\pm$6 \% in the calibrations of the SRM in the HER and LER respectively, 
(2) 2 \% in the HER due to uncertainty in the location of the source point. 
In the LER, this error is much less than 1 \% because the difference of the betatron function between the both ends of the bending magnets where the synchrotron light is emitted is very small.
(3) Errors in measurements of betatron functions at the source points are estimated to be about $\pm$5 \%. 
The resultant values are 1.000 $\pm$ 0.045 and 0.971 $\pm$ 0.060 in the HER and the LER, respectively.

\section{X-Y COUPLING}\
We estimated x-y coupling in each ring using the corrected vertical beam size of the interaction point (IP), which is transferred from the source point of SRM, as shown in Table~\ref{xy}.
The design values are used for the horizontal emittances, because the horizontal beam size measurement cannot be checked on by the method using the iSize bumps. 
The horizontal beam size should be also checked by the other method in the future.
The obtained coupling parameters are much smaller than those of colliding beams which are about 3 \%.
However, the beam-beam simulations predict that there is still room for improvement to reduce the coupling of single beam to obtain the higher luminosity.
\begin{table}[h]
\begin{center}
\caption{The x-y couplings calculated using the corrected vertical beam sizes}
\vskip1mm
\begin{tabular}{|c||c*{3}{|c}||c|}
\hline
                  &  $\sigma_Y^{\ast}$   & $\beta_Y^{\ast}$   & $\varepsilon_Y$    & $\varepsilon_X$ & $\kappa$   \\ 
                 &  ($\mu$m)                   & (mm)                        & (nm)                         & (nm)                       &( \%)            \\ \hline
HER        & 1.32                               & 5.9                            & 0.30                         & 24                          & 1.2              \\
LER         & 1.21                              & 5.9                            & 0.25                         & 18                          & 1.4              \\
\hline
\end{tabular}
\label{xy}
\end{center}
\end{table}
\section{CONCLUSION}
We have presented a check on the calibration factors for the SRMs which are always used for measuring the transverse beam sizes in KEKB.
It is useful for checking on the calibration to compare the vertical beam sizes measured by SRM with those calculated by SAD computer code at various iSize bump heights.
The results are consistent with the SRM's calibration factors within errors.
The x-y couplings of both rings are calculated with the calibrated vertical beam sizes.
The reliable beam size measurement with the cross-check of the calibration factor is very important, since an achievable luminosity is dependent on the single-beam emittances.

\end{document}